\documentstyle[12pt] {article}
\newcommand{\beq}{\begin{equation}}
\newcommand{\eeq}{\end{equation}}
\newcommand{\beqa}{\begin{eqnarray}}
\newcommand{\eeqa}{\end{eqnarray}}
\newcommand{\ba}{\begin{array}}
\newcommand{\ea}{\end{array}}
 
\begin{document}

\def \mtil{{\tilde m}}

\begin{center}
{\bf SPECTRAL DISTRIBUTION STUDIES OF fp SHELL NUCLEI \\
WITH MODIFIED KUO--BROWN INTERACTION} \\
\vskip 0.5 truecm
{\bf K. Kar} and {\bf S. Sarkar} \\
Saha Institute of Nuclear Physics \\
Block - AF, Sector I, Bidhan Nagar \\
Calcutta 700 064, India \\
\vskip 0.5 truecm
{\bf J.M.G. Gomez} \\
Departamento de Fisica Atomica, Molecular y Nuclear \\
Facultad de Ciencias Fisicas, Universidad Complutense de Madrid \\
E-28040 Madrid, Spain \\
\vskip 0.5 truecm
{\bf V.R. Manfredi} \\
Dipartimento di Fisica "G. Galilei", Universit\`a di Padova \\
Istituto Nazionale di Fisica Nucleare - Sezione di Padova \\
via Marzolo 8, I-35131, Padova, Italy \\
\vskip 0.5 truecm
{\bf L. Salasnich} \\
Dipartimento di Matematica Pura ed Applicata \\
Universit\`a di Padova, via Belzoni 7, I-35131 Padova, Italy \\
Istituto Nazionale di Fisica Nucleare, Sezione di Padova \\
via Marzolo 8, I-35131, Padova, Italy \\
\end{center}

\newpage

\begin{center}
{\bf ABSTRACT}
\end{center}

\vskip 0.5 truecm


\newpage

\section{INTRODUCTION}

In  recent  years there has been substantial progress in the
application of shell model to study nuclear  structure.  Full  fp
shell calculations involving valence particles in all four orbits
$f_{7/2}$, $f_{5/2}$, $p_{3/2}$ and $p_{1/2}$ have been 
successfully completed [1,2]. New realistic interactions in 
the fp shell have been suggested for better agreement with 
experimental results for the binding energies, low-lying spectra 
and excitation strengths. These studies are being carried further
to understand many other microscopic features of the nuclei in
this  region. Some of these nuclei are also important in
astrophysics, in particular for pre--supernova stellar evolution
[3,4] and  r- and s-process nucleosynthesis. But for 
astrophysical purposes, one often finds that average properties, 
like smoothed level densities and averaged strength functions, are
adequate. Here, results of statistical models of nuclear structure
are useful. Spectral distribution theory [5,6] is a 
theory  which, given enough valence particles in large spaces, is
able to give statistically smoothed average shell model values 
for the physical quantities of interest.

In this paper we shall be concerned with the applications of 
the spectral  distribution theory to the fp shell. 
In all earlier such studies the major
uncertainties arose from the interaction used and none of the
interactions  used  could  give  results in good agreement with
observed values over the whole lower/upper half of the shell. But
recently, shell model studies of A=48 nuclei [1] as well as  some
other heavier ones [7] in the lower half of the shell indicate
that a minimally modified Kuo-Brown interaction (KB3) is able  to
reproduce  successfully experimental binding energies, excitation
spectra  and  transition  strengths.  Then  the   question   that
naturally  arises  is  how  well  does  the spectral distribution
studies do with this interaction in the  lower  half  of  the  fp
shell.  In  this  work, we compare the predictions of the spectral
distribution methods with experimental and  shell  model  values.
Similar  studies were carried out in the sd shell [8] after
the spectacular success of shell model results with  universal-sd
interaction [9].

In spectral distribution theory one produces 
smoothed fluctuation free forms for  the  density  of  states  by
distributing  $m$  fermions over $N$ single particle states which
go asymptotically to Gaussians.  One  is  also  able  to  provide
average  expectation values of operators as polynomial expansions
in terms of energy of the initial space. The partitioning of  the
full  shell  model space into configurations and the use of a Gaussian
form  for  the  density  in  each  configuration   improves   the
predictability  of the position of discrete states as well as the
expectation values of operators and other relevant quantities. In
predicting the binding energy through spectral distributions,  one
often  uses  the  experimental  spectra  and  does integration of
Gaussians up to  an  excited  state,  and  then  subtracts  out  the
excitation  energy  to  reduce  the  inaccuracy  coming  from the
integration  procedure. The other   correction   one   should
incorporate   is   the  small  but  non-zero skewness  and  excess
$(\gamma_1,\gamma_2)$ of the distribution coming from  large  but
finite  shell  model  spaces.  All  earlier  studies  of spectral
distributions in fp shell used the excited state  correction,  but
in    this   paper   for   the   first   time   we   incorporate
$(\gamma_1,\gamma_2)$  corrections  for  fp   shell   nuclei   in
evaluating   binding   energies, excitation   spectra   and  orbit
occupation probabilities. A comparison with  experimental  values
shows  the  importance of taking into account this deviation from
Gaussians  in  improving  predictions.  This  feature  was   also
observed in the sd shell comparisons.

\section{FORMALISM}

In  the  shell  model  space  of  m  particles
(called the scalar space) the density  of  states  goes  towards  a
Gaussian,  which  needs  two  quantities  the  centroid  $E_c(m)$
[$=\langle  H  \rangle^m $]  and  the  width  $\sigma^2  (m)$
[$=\langle \tilde H^2 \rangle^m = \langle(H-\langle H\rangle^m)^2\rangle^m$ ] 
to be specified. 
Here the m-particle  average is given by $\langle H \rangle^m = Tr~ H/d(m)$ 
where $Tr~H$ is the trace of the Hamiltonian operator $H$ and d(m) 
is the dimension  of the shell model space. 
The skewness and excess are then given by
$$
\gamma_1(m) = \langle {\tilde H}^3 \rangle^m/\sigma^3 (m) 
$$
$$
\gamma_2(m) = (\langle {\tilde H}^4 \rangle^m/\sigma^4 (m)) -3 . \eqno (1) 
$$

Given    the    (1+2)-body   realistic   Hamiltonians,   spectral
distribution theory expresses the m-particle averages in terms of
averaged 1-  and  2-body  matrix  elements  and  propagators(which
involve  powers  of m $ [6]$).  For  application to real nuclei, one
needs to work in (m,T) spaces where T stands for the  isospin  of
the  m-particle state. Spectral distributions also demonstrate
the Gaussian forms for the (m,T)  density  of  states  and  give 
extensions  of  the  propagation  results  for  (m,T)  as well as
$(\mtil,T)$ spaces [5]. $(\mtil,T)$ stands for 
configuration-isospin space where $\mtil=m_1,m_2,...,m_l$ are the
particles  in  $l$  orbits.  The  ground  state  energy  $\bar{E}_g$ is
evaluated by a  procedure  due  to  Ratcliff [10] where  one
inverts the equation
$$
\sum_\mtil  \int_{-\infty}^{\bar{E}_g} I_{\mtil  T}  (E) dE =d_0/2
\eqno (2)
$$
to get $\bar{E}_g$ ($d_0$ is the degeneracy of the ground state ). 
Here $I_{\mtil,T}(E)=d(\mtil,T) \rho(\mtil,T)$. 
The  expression  for  the  Gaussian  density  of  states  in
$(\mtil,T)$ space is
$$
\rho  (\mtil,T)  = {1\over \sqrt(2\pi)\sigma(\mtil, T)} exp \big
[-{1\over2} (E -  E_c(\mtil,T))^2/\sigma^2(\mtil,T)\big  ]  \eqno (3)
$$
To incorporate
the  $(\gamma_1,\gamma_2)$  correction  we  take  recourse  to  the
Cornish-Fisher expansion $[6]$. In this expansion  one  transforms  the
variable  $x$  in $\rho(x)$ by a series expansion onto a variable
$y$ so that the density in $y$ is a Gaussian $\rho_G (y)$. 
Then for densities in $x$ and $y$ both with zero centroid and unit width 
one gets, including the $(\gamma_1,\gamma_2)$ corrections
$$
y = x  -  {\gamma_1 \over 6} (x^2-1)  +  [-{\gamma_2 \over 24} (x^3-3x)  +
{\gamma_1^2\over 36}  (4x^3-7x)] \eqno (4)
$$
and conversely
$$
x = y  +  {\gamma_1 \over 6}(y^2-1)  +  [{\gamma_2 \over 24}(y^3-3y)  -
{\gamma_1^2 \over 36} (2y^3-5y)]
\eqno(5)
$$
so that  $\rho(x) = \rho_G(y){dy\over dx} $. The orbit occupation
probability for orbit s in the m - particle space is given simply
by   $$   n_s(E)   =    \sum_\mtil    {I_{\mtil    T}(E)    \over
I_{mT}(E)}[m_s(\mtil)]   \eqno(6)$$  This gives a simple 
dependence of the occupation probability on the energy E [11].

In spectral distribution theory, for comparison of different 
operators, an important quantity is the correlation coefficient
between two operators $G$ and $H$ defined by
$$
\zeta_{G-H} = {\langle (G-\langle G \rangle)(H-\langle H \rangle) \rangle^m  
\over \sigma_{G} (m) \sigma_{H} (m)} 
$$
where the $m$-particle trace $<\tilde G \tilde H>^m$ is calculated 
using propagation techniques and $\sigma_{G} (m)$ ($\sigma_{H} (m)$) 
are the widths of $G$($H$) in the m-particle space. 
The extension to (m,T) space also is easily carried out [5].

Our  spectral  distribution codes as yet can calculate up to third
moments in (m,T) spaces exactly.  The  fourth  moment  of  2-body
operators  can  be  calculated  only  in  scalar  spaces.  So for
$\gamma_2(m,T)$ we first make  an  approximation  $\gamma_2(m,T)=
\gamma_2(m)$  to calculate the binding energies and spectra; then
we  improve  this  approximation  by  using  a   phenomenological
correction  term involving the two scalars of isospin space n and
$T^2$ and write $\gamma_2(n,T)= 0.04n - 0.04 T^2$. The correction
coming from $\gamma_1$ in the energy is small (a few percent); so
changing the $\gamma_1$ from its scalar to exact  (m,  T)  values
hardly  makes  any  change in the corrected energy. Therefore we keep the
scalar value for our calculation.  

\section{RESULTS AND DISCUSSIONS}

In Table 1 we compare the 
predictions  for  a  number of nuclei in the lower half of the fp
shell with the experimental binding energies  (with  the  Coulomb
contribution removed from it). Table 1 also gives the predictions
of Haq and Parikh [12] using configuration isospin moment with excited
state  correction  using  MHW2 interaction.  We find that our
procedure  gives  substantially better  agreement  with   experimental   values
compared   to  earlier  SDM  applications  particularly for  nuclei
with large ground state isospin values.
The average and rms deviation  of  the  corrected
(column  C) binding energies from the experimental values are 0.15 and 1.49 MeV 
respectively. Kota and Potbhare using SDM with excited  state  corrections
with a phenomenological term involving neutron and proton numbers
got  the RMS deviation as 5.59, 2.19, 5.79, 8.39 and 3.60 MeV for
KB, MHW2 KB10, bare and MWH interactions respectively [13]. So
we see that incorporating the corrections in binding energies due
to non-zero $(\gamma_1,\gamma_2)$ values make substantial improvements 
compared to other methods using spectral distributions. Bearing 
in mind that fluctuations are of the order of 1 MeV, we find that
this is a very satisfactory procedure.

To  understand how the present interaction KB3 differs from
earlier interactions, like MHW2 which was also derived  from  Kuo-
Brown  interaction,  we display in Table 2 the centroid and width
of the two interactions  and  their  correlation  coefficient  in
scalar  -  isospin  fp  spaces.  These  quantities, as one number
estimates, give the overall behaviour of the interactions. 
The interaction KB3 [1] 
is obtained by subtracting out 300 keV for J=1,3 with T=0 and 200 keV 
for J=2 with T=1 from the diagonal matrix elements of the $f_{7/2}$ 
orbit of KB1. KB1 in turn is obtained by modifying some diagonal 
elements of the original Kuo-Brown interaction [1]. 
The centroids of KB3 and MHW2 are found to differ by up to 6 MeV
for  particle number ranging from 6 to 16. The width of MHW2 is
seen to be consistently smaller than KB3 by a few percent, but as the
correlation coefficient has the centroid subtracted and the widths divided
out it has values very close to one for all particle numbers and isospins.

The  procedure  for  calculating  the  energy  of  states  can be
extended to excited states also. In Fig. 1 we  compare for the nuclei $^{46}Ti$
 and $^{48}Sc$ the calculated excitation spectrum with observed spectrum as well
as   shell   model   ones  (for $^{48}Sc$) obtained using the same KB3 
interaction. The  spectral  distribution gives a globally averaged spacing and
as a result does not reproduce well the clustering of  states  at
low excitation energies for the  odd-odd nucleus $ ^{48}Sc$.
 Also in spectral distribution studies the spin
sequence is assumed to locate each excited state. But we see that allowing
for fluctuations of individual levels, the overall spectrum is reproduced
quite well by spectral distributions for both the examples.

Finally in Table 3 we give the ground state occupation
probabilities of the four orbits $f_{7/2}$, $p_{3/2}$,  $f_{5/2}$
and  $p_{1/2}$  by our method. 
As is well known, the occupation probabilities of an orbit
is related to the sum rule of stripping and pick-up strengths. But the
analysis of the experimental results are not done for too many nuclei
for direct comparison and even 
 the data available have large uncertainties. We
quote the experimental ground state occupancies given in Kota and
Potbhare [12] for nuclei $^{46}Ti$(T=1), $^{48}Ti$(T=2), $^{52}Cr$(T=2) 
and $^{56}Fe$(T=2). For $^{48}Ti$ and $^{52}Cr$ the $f_{7/2}$
occupancies calculated by us agree reasonably well with experiments, 
but for $^{46}Ti$ and $^{56}Fe$ our values are higher. One feels the
need for a more systematic analysis of present pick-up/stripping
experiments and to perform further experiments for a more detailed 
comparison. The occupancies are quite useful for the estimation of Gamow-Teller
sum rule strengths for $\beta^-$ and $\beta^+$ decays [3,14].

\section{CONCLUSIONS}

In conclusion, we stress that spectral distribution studies using 
corrections derived from a departure from Gaussians for the density of states
through the 3rd and 4th moments of the Hamiltonian are quite successful in
predicting binding energies, excitation spectra etc. These studies should be
extended to the calculation of sum rules and transition strength distributions
for different excitation operators and also to the upper half of the fp
shell.

\section*{ACKNOWLEDGMENT}

We would like to thank V.K.B. Kota for letting us use some of his computer
programmes.

\newpage

{\bf REFERENCES}
\vskip 0.5 truecm

[1] E. Caurier, A.P. Zuker, A.Poves and G. Martinez - Pinedo,
Phys. Rev. C50 (1994) 225

[2] K. Langanke, D.J. Dean, P.B. Padha, Y. Alhassid and S.E.
Koonin, Phys. Rev. C52 (1995) 718

[3] K. Kar, A.Ray and S. Sarkar, Astrophys. Jour. 434 (1994) 662

[4] M.B. Aufderheide, G.E. Brown, T.T.S. Kuo, D.B. Stout and P.
Vogel, Astrophys. Jour. 362 (1990) 241

[5] J.P. Draayer, J.B. French and S.S.M. Wong, Ann. Phys. (NY)
106 (1977) 472; J.B. French and V.K.B. Kota, Ann. Rev. Nucl. Part. Sci.
32 (1982) 35

[6] V.K.B. Kota and K.Kar, Lab. Rep., UR-1058 (1988) Univ
Rochester and Pramana 32 (1989) 647

[7] E. Caurier, G. Martinez - Pinedo, A. Poves and A.P. Zuker,
 Phys. Rev. C52 (1995) R1736

[8] S. Sarkar, K. Kar and V.K.B. Kota, Phys. Rev. C36 (1987) 2700

[9] B.H. Wildenthal, Prog. Part. Nucl. Phys. 11 (1984) 5

[10] K.F. Ratcliff, Phys. Rev. C3 (1971) 117

[11] V. Potbhare and S.P. Pandya, Nucl. Phys. A256 (1976) 253

[12] R.U. Haq and J.C. Parikh, Nucl. Phys. A273 (1976) 410

[13] V.K.B. Kota and V. Potbhare, Nucl. Phys. A 331 (1979) 93

[14] G.M.Fuller, W.A.Fowler and M.J.Newman, Astrophys. Jour. 252 (1982)
 715; S.Sarkar and K.Kar, Jour. Phys. G14 (1988)L123.

\newpage

\begin{center} 
{\bf Table 1}
\end{center}

Binding energies(BE) of nuclei in the lower half of fp shell by spectral
distribution methods(SDM) with KB3 interaction compared to experimental
binding energies. Column $\bar A$ gives BE by Ratcliff procedure, Column
$\bar B$ and $\bar C$ by Ratcliff procedure with ($\gamma_1,\gamma_2$)
corrections with $\gamma_1(m)$,$\gamma_2(m)$ and $\gamma_1(m,T)$,
$\gamma_2(m,T)$ values respectively. 

$$\halign{&\hfil\  # \ \hfil \ \cr
\noalign{\hrule}\cr
\multispan{2}\hfil Nucleus \hfil&Expt 
BE&\multispan{3}\hfil BE by SDM\hfil&BE by SDM\cr
    & &(in MeV)&\multispan {3} \hfil (in MeV) \hfil & (in MeV) \cr
A  & Z     &     &$\bar A  $ &$ \bar B$&$ \bar C$& (Haq \& Parikh) \cr
\noalign{\hrule}\cr
46&20&-56.79&-60.73&-56.70&-56.33&-58.93\cr
46&21&-62.95&-66.36&-62.89&-62.89&-64.94\cr
46&22&-71.49&-74.15&-69.08&-69.43&-70.53\cr
46&23&-&-75.82&-71.01&-71.35&-\cr
48&20&-73.84&-77.70&-72.60&-72.31&-76.72\cr
48&21&-81.71&-85.88&-81.32&-80.96&-81.56\cr
48&22&-92.34&-98.16&-90.76&-91.05&-93.08\cr
48&23&-94.94&-100.39&-94.15&-94.88&-96.36\cr
48&24&-101.16&-104.96&-97.48&-98.64&-98.99\cr
52&20&-95.18&-101.76&-97.05&-93.82&-100.61\cr
52&21&-109.43&-120.35&-110.88&-108.78&-112.87\cr
52&22&-126.02&-135.36&-125.96&-124.26&-127.37\cr
52&23&-134.29&-144.70&-134.94&-134.94&-138.23\cr
52&24&-145.63&-156.98&-143.71&-145.50&-146.25\cr
52&25&-148.41&-158.37&-146.97&-149.54&-148.88\cr
52&26&-154.22&-164.13&-150.49&-154.18&-153.18\cr
\noalign{\hrule}\cr}$$

\newpage 

{\bf Table 1 (contd.)}
$$\halign{&\hfil\  # \ \hfil \ \cr
\noalign{\hrule}\cr
\multispan{2}\hfil Nucleus \hfil&Expt BE&\multispan{3}
\hfil BE by SDM\hfil&BE by SDM\cr
    & &(in MeV)&\multispan {3} \hfil (in MeV) \hfil & (in MeV) \cr
A  & Z     &     &$\bar A  $ &$ \bar B$&$ \bar C$& (Haq \& Parikh) \cr
\noalign{\hrule}\cr
56&20&-108.41&-112.25&-110.50&-108.79&-112.03\cr
56&21&-126.95&-132.93&-130.10&-127.36&-133.27\cr
56&22&-148.26&-157.94&-150.99&-146.71&-152.08\cr
56&24&-177.96&-192.22&-177.24&-178.03&-180.18\cr
56&25&-187.17&-201.80&-188.21&-189.48&-188.05\cr
56&26&-198.93&-214.49&-196.62&-200.79&-198.63\cr
56&27&-202.72&-216.51&-200.29&-205.60&-201.41\cr
56&28&-208.66&-222.21&-203.46&-210.46&-207.29\cr
\noalign{\hrule}\cr}$$

\newpage

\begin{center} 
{\bf Table 2}
\end{center}

Centroids, widths and the correlation coefficient for the interactions
modified  Kuo-Brown (KB3) and MHW2.

$$\halign{&\hfil\  # \ \hfil \ \cr
\noalign{\hrule}\cr
Number&&\multispan{2}\hfil KB3  \hfil&\multispan{2} \hfil MHW2\hfil
&Correlation \cr
of    &Iso- &\multispan {2} \hfil     \hfil &&&
Coefficient between \cr
valence&-spin&Centroid&Width&Centroid&Width&KB3\&MHW2 \cr
particles&     &(MeV)&(MeV)&(MeV)&(MeV) \cr
\noalign{\hrule}\cr
6&0&-42.04&8.33&-40.92&7.95&0.999 \cr
 &1&-40.63&8.06&-39.72&7.72&0.999 \cr
 &2&-37.81&7.49&-37.31&7.24&0.999 \cr
 &3&-33.59&6.57&-33.70&6.48&1.000 \cr
 \cr
8&0&-58.84&9.85&-57.01&9.32&0.998 \cr
 &1&-57.44&9.61&-55.81&9.11&0.998 \cr
 &2&-54.62&9.12&-53.40&8.69&0.998 \cr
 &3&-50.40&8.35&-49.79&8.03&0.998 \cr
 &4&-44.77&7.23&-44.98&7.10&1.000 \cr
 \cr
12&0&-96.66&12.26&-92.86&11.42&0.997 \cr
  &1&-95.24&12.06&-91.66&11.23&0.997 \cr
  &2&-92.43&11.64&-89.25&10.85&0.997 \cr
  &3&-88.21&11.01&-85.25&10.28&0.997 \cr
  &4&-82.58&10.13&-80.83&9.51&0.997 \cr
  &5&-75.54&8.97&-74.82&8.51&0.998 \cr
  &6&-67.09&7.46&-67.60&7.24&0.999 \cr
  \cr
16&0&-140.06&13.89&-133.61&12.73&0.996 \cr
  &1&-138.65&13.71&-132.41&12.55&0.995 \cr
  &2&-135.84&13.33&-130.00&12.20&0.995 \cr
  &3&-131.61&12.76&-126.39&11.66&0.995 \cr
  &4&-125.98&11.98&-121.58&10.94&0.995 \cr
  &5&-118.94&10.98&-115.57&10.03&0.994 \cr
  &6&-110.50&9.72&-108.35&8.90&0.995 \cr
  &7&-100.64&8.14&-99.92&7.52&0.996 \cr
  &8&-89.38&6.06&-90.30&5.80&0.999 \cr
\noalign{\hrule}\cr}$$

\newpage

\begin{center} 
{\bf Table 3}
\end{center}

Calculated occupancies for the fp - shell nuclei. The values in
parenthesis are from experimental data$^{\dagger})$ obtained by adding
neutron and proton occupancies.

$$\halign{&\hfil\  # \ \hfil \ \cr
\noalign{\hrule}\cr
Atomic&Number of&&\multispan{4}\hfil Occupancy  \hfil \cr
Number&valence&Isospin&$f_{7/2}$&$f_{5/2}$&$p_{3/2}$&$p_{1/2}$
\cr
&particles\cr
\noalign{\hrule}\cr
46&6&0&5.77&0.03&0.18&0.02 \cr
  & &1&5.79&0.01&0.18&0.02 \cr
  & & &(4.89)&(0.23)&(0.88)&(0.00) \cr
  & &2&5.66&0.01&0.30&0.03 \cr
  & &3&5.58&0.00&0.39&0.03 \cr
  \cr
48&8&0&7.51&0.09&0.34&0.06 \cr
  & &1&7.39&0.09&0.44&0.08 \cr
  & &2&7.38&0.06&0.48&0.08 \cr
  & & &(7.08)&(0.14)&(0.78)&(0.14) \cr
  & &3&6.89&0.08&0.88&0.15 \cr
  & &4&6.62&0.06&1.14&0.18 \cr
\cr
52&12&0&10.58&0.36&0.84&0.22 \cr
  &  &1&10.28&0.40&1.04&0.28 \cr
  &  &2&10.13&0.35&1.21&0.31 \cr
  & & &(9.98)&(0.06)&(1.96)&(0.00) \cr
  &  &3&9.50&0.42&1.65&0.43 \cr
  &  &4&8.97&0.42&2.07&0.54 \cr
  &  &5&8.18&0.57&2.52&0.73 \cr
  &  &6&7.53&0.66&2.90&0.91 \cr
\cr
56&16&0&12.96&0.78&1.74&0.52 \cr
  &  &1&12.60&0.85&1.96&0.59 \cr
  &  &2&12.31&0.85&2.19&0.65 \cr
  & & &(10.95)&(1.94)&(2.74)&(0.37) \cr
  &  &3&11.62&1.00&2.58&0.80 \cr
  &  &4&11.02&1.09&2.94&0.95 \cr
  &  &6&9.58&1.50&3.58&1.34 \cr
  &  &7&8.79&1.90&3.79&1.52 \cr
  &  &8&8.00&1.43&3.95&1.68 \cr
\cr
\noalign{\hrule}\cr}$$
\leftline {$ ^{\dagger)}$ Ref. [13] } 

\newpage

\begin{center} 
{\bf Figure Caption}
\end{center}

            {\bf Figure 1}: The excitation spectrum of $ ^{48}Sc$ and $
            ^{46}Ti$ calculated by spectral distributions (SDM) compared
            with the experimental and shell model (for $ ^{48}Sc$)
            spectra. The interaction used for the SDM and shell model
            is the modified Kuo-Brown (KB3). 

\end{document}